\documentstyle[preprint,aps,pra]{revtex}
\begin{document}
\draft
\title{Radiation field quantization in a nonlinear dielectric
with dispersion and absorption}
\author{E. Schmidt, L. Kn\"oll, and D.-G. Welsch}
\address{Friedrich-Schiller-Universit\"{a}t Jena,\\ Theoretisch-Physikalisches\\
Institut\\ Max-Wien-Platz 1, D-07743\\ Jena,\\ Germany }
\date{July 31, 1996}
%\date{\today}
\maketitle

\begin{abstract}
The problem of
quantizing the radiation field inside a nonlinear dielectric is studied.
Based on the quantization of radiation in a linear dielectric which includes
absorption and dispersion, we extend the theory in order to treat also
nonlinear optical processes. We derive propagation equations in space and
time for the quantized radiation field including the effects of linear
absorption and dispersion as well as nonlinear optical effects. As a special
case we derive the propagation equation of a narrow-frequency band light
pulse in a Kerr medium.
\end{abstract}

%\date{July 19, 1996}

\narrowtext

\section{Introduction}

\label{s1} Non-classical properties of quantum light pulses propagating
through nonlinear optical media are of increasing interest. E.g. quantum
solitons in fibers may in future play an important role in communication
technology \cite{Drummond.Nature.1993}. But the proper consideration of
quantum effects as well as changes of quantum properties of light pulses
propagating through nonlinear optical media requires an adequate description
of the quantized radiation field in such media. In addition to the nonlinear
optical properties of the dielectric under consideration, also effects of
linear absorption and dispersion must be included. As it is well known the
existence of solitons in fibers with a Kerr-like nonlinearity requires also
the dispersion of the group velocity of the wave \cite
{Hasegawa.1989,Akhmanov.1992}. On the other hand non-classical phenomena
such as squeezing are very sensitive to degradation effects as for instance
absorption of radiation. Therefore it is necessary, to develop a theory of
the quantized radiation field in dielectrics which considers both nonlinear
optical effects and dispersion as well as absorption effects. The
quantization of the radiation field in dielectrics has been performed in the
past in different ways. Some of this treatments are presented in
\cite
{Hopf.1958,Pantell.1969,Marcuse.1980,Yariv.1989,%
Abram.1987,Kn.Vo.We.1987,Glauber.Lewenstein.1991,%
Hillery.1984,Drummond.1987,Lai.Haus.1989a,Lai.Haus.1989b,%
Drummond.1990,Huttner.Barnett.1992,Ho.Kumar.1993,%
Gruner.Welsch.1995,Matloob.1995,Gruner.Welsch.1996,Hillery.1996}.
The quantization of the electromagnetic field in linear, lossless, and
dispersionless dielectric media has been developed in
\cite
{Pantell.1969,Marcuse.1980,Yariv.1989,Abram.1987,%
Kn.Vo.We.1987,Glauber.Lewenstein.1991}.
The case of lossless and dispersionless nonlinear dielectrics has been
treated by Hillery and Mlodinov in \cite{Hillery.1984}. Squeezing of
solitons in dispersive nonlinear dielectrics was successfully explained in
\cite{Drummond.1987}. A canonical quantization schema for a general
dispersive nonlinear dielectric has been presented by Drummond \cite
{Drummond.1990}. The methods of the phase-space formulation of the equations
for quantum optical pulse propagation in nonlinear and dispersive
single-mode fibers are discussed in \cite{Carter.1995}. A very interesting
approach to the quantization of the radiation field in a dielectric composed
of two-level atoms has been recently proposed in \cite{Hillery.1996}.

Huttner and Barnett \cite{Huttner.Barnett.1992} presented a canonical
quantization scheme for the radiation field in a linear dielectric,
employing the Hopfield model of the dielectric \cite{Hopf.1958}. It is worth
noting that in this approach both the dispersion and the absorption by the
medium are taken into account in a quantum-mechanically consistent way.
A phenomenological generalization of this theory was given in \cite
{Gruner.Welsch.1995,Matloob.1995,Gruner.Welsch.1996,Gruner.Welsch.io},
and the theory has been
applied to the propagation of quantum pulses in \cite{Schmidt.1996}. In this
paper we develop a theory of the quantized radiation field in dielectrics
which includes both nonlinear optical effects and dispersion as well as
absorption effects. We employ the quantization scheme developed by Huttner
and Barnett \cite{Huttner.Barnett.1992}. This quantization schema is based
on a microscopic model of Hopfield type \cite{Hopf.1958}, where the
dielectric medium is represented by a polarization field $X\left( x\right)$.
We extend this quantization scheme by introducing nonlinear polarization
field terms into the Hamiltonian. In reality the material system is not a
linear one, and these additional terms describe the intrinsic nonlinear
properties of the dielectric which can give rise to nonlinear optical
effects. In Sec.\ref{s2} we review the main results of the quantization
scheme for a linear dielectric described in \cite
{Huttner.Barnett.1992,Gruner.Welsch.1996}. In Sec.\ref{s3} the theory is
extended to nonlinear dielectrics, and the quantization schema in this case
is described, including linear dispersion and absorption. A propagation
equation for a narrow-frequency band quantum pulse is derived in Sec.\ref{s4}
and the special case of a Kerr-like nonlinearity is considered in more
detail. A summary is given in Sec.\ref{s5}.

\section{Linear medium}

\label{s2}

In this section we review some results of the quantization of the
electromagnetic field in a dispersive and absorptive linear dielectric. We
employ the quantization scheme developed by Huttner and Barnett \cite
{Huttner.Barnett.1992} and extended in \cite{Gruner.Welsch.1996}. This
quantization schema is based on a microscopic model of Hopfield type \cite
{Hopf.1958} where the dielectric medium is represented by a polarization
field $X\left( x\right) $, which is coupled to a reservoir composed of a
continuum of harmonic oscillators $Y\left( x,\omega \right) $ to allow for
absorption ($0<\omega <\infty $, the continuum of bath modes at a fixed
space point $x$ is labeled by the parameter $\omega $). For convenience we
consider only a one-dimensional model of the dielectric, the propagation
direction of the radiation field is the x-direction and the vector potential
$A\left( x\right) $ is linear polarized in the z-direction. The
electromagnetic field is coupled to the polarization field in the minimal
coupling version, and the resulting Hamiltonian can be diagonalized
introducing bosonic operators $\hat f\left( x,\omega \right) $ and $%
\hat f^{\dagger }\left( x,\omega \right) $, which describe
polariton-like excitations inside the dielectric \cite
{Huttner.Barnett.1992,Gruner.Welsch.1996}. After diagonalization the
Hamiltonian of the transverse electromagnetic field interacting with the
linear dielectric can be written as
\begin{equation}
\label{def.HL.01}{\bf \hat H}=\hbar {\hat {{\cal H}}}_L=\hbar
\int\limits_{-\infty }^\infty dx\ \hat H_L\left( x\right) ,
\end{equation}
where
\begin{equation}
\label{conn.HL.wff}\hat H_L\left( x\right) =\int\limits_0^\infty d\omega
\ \omega \ \hat f^{\dagger }\left( x,\omega \right) \hat f\left(
x,\omega \right) .
\end{equation}
The creation and destruction operators $\hat f^{\dagger }\left( x,\omega
\right) $ and $\hat f\left( x,\omega \right) $ satisfy the familiar
bosonic equal time commutation relations
\begin{equation}
\label{com.rel.ff}\left[ \hat f\left( x,\omega \right) ,\hat f%
^{\dagger }\left( x^{\prime },\omega ^{\prime }\right) \right] =\delta
\left( x-x^{\prime }\right) \ \delta \left( \omega -\omega ^{\prime }\right)
,
\end{equation}
\begin{equation}
\label{com.rel.fff}\left[ \hat f\left( x,\omega \right) ,\hat f%
\left( x^{\prime },\omega ^{\prime }\right) \right] =0.
\end{equation}
The operators of the electromagnetic field, e.g. the vector potential $%
\hat A\left( x\right) $ and the electric field strength $\hat E%
\left( x\right) $ as well as the polarization field operator $\hat X%
\left( x\right) $ and the bath operators $\hat Y\left( x,\omega \right) $
may be written in terms of the bosonic operators $\hat f^{\dagger
}\left( x,\omega \right) $ and $\hat f\left( x,\omega \right) $ as
follows
\begin{equation}
\label{conn.A.int.dw.A}\hat A\left( x\right) =\hat A^{(+)}\left(
x\right) +\hat A^{(-)}\left( x\right) =\int\limits_0^\infty d\omega
\left[ \hat A\left( x,\omega \right) +\hat A^{\dagger }\left(
x,\omega \right) \right] ,
\end{equation}
\begin{equation}
\label{conn.E.int.dw.E}\hat E\left( x\right) =\hat E^{(+)}\left(
x\right) +\hat E^{(-)}\left( x\right) =\int\limits_0^\infty d\omega
\left[ \hat E\left( x,\omega \right) +\hat E^{\dagger }\left(
x,\omega \right) \right] ,
\end{equation}
\begin{equation}
\label{conn.X.int.dw.X}\hat X\left( x\right) =\hat X^{(+)}\left(
x\right) +\hat X^{(-)}\left( x\right) =\int\limits_0^\infty d\omega
\left[ \hat X\left( x,\omega \right) +\hat X^{\dagger }\left(
x,\omega \right) \right] ,
\end{equation}
where
\begin{equation}
\label{conn.Aw.int.dw.GA.f}\hat A\left( x,\omega \right)
=\int\limits_{-\infty }^\infty dx^{\prime }\ G_{{\rm A}}\left( x,x^{\prime
},\omega \right) \ \hat f\left( x^{\prime },\omega \right) ,
\end{equation}
\begin{equation}
\label{conn.Ew.Aw}\hat E\left( x,\omega \right) =i\omega \hat A%
\left( x,\omega \right) ,
\end{equation}
\begin{equation}
\label{conn.Xw.Ew.f}\frac \rho {\epsilon _0}\hat X\left( x,\omega
\right) =\left( \varepsilon \left( \omega \right) -1\right) \ \hat E%
\left( x,\omega \right) -2i\alpha c
\sqrt{\varepsilon _{{\rm i}}\left( \omega \right)}\ \hat f%
\left( x,\omega \right) .
\end{equation}
The Green function $G_{{\rm A}}\left( x,x^{\prime },\omega \right) $ is
given by
\begin{equation}
\label{eq.GA}G_{{\rm A}}\left( x,x^{\prime },\omega \right) =-i\alpha \sqrt{%
\frac{\varepsilon _{{\rm i}}\left( \omega \right) }{\varepsilon \left(
\omega \right) }}\ e^{ik\left( \omega \right) \left| x-x^{\prime }\right| },
\end{equation}
where $\varepsilon \left( \omega \right) =\varepsilon _{{\rm r}}\left(
\omega \right) +i\varepsilon _{{\rm i}}\left( \omega \right) $ denotes the
complex permittivity of the dielectric. The complex wave number $k\left(
\omega \right) $ is related to the complex permittivity $\varepsilon \left(
\omega \right) $ and the complex index of refraction $n\left( \omega \right)
$ by
\begin{equation}
\label{def.k}k\left( \omega \right) =\frac \omega c\ \sqrt{\varepsilon
\left( \omega \right) }=\frac \omega c\ n\left( \omega \right) =\frac \omega
c\ \left( n_{{\rm r}}\left( \omega \right) +n_{{\rm i}}\left( \omega \right)
\right) =k_{{\rm r}}+ik_{{\rm i}}.
\end{equation}
A normalization constant $\alpha $ has been introduced for convenience
\begin{equation}
\label{def.alpha}\alpha =\sqrt{\frac \hbar {4\pi c^2\epsilon _0{\cal A}}},
\end{equation}
where $\epsilon _0$ is vacuum permittivity constant, and ${\cal A}$ is the
normalization area perpendicular to the $x$ direction. The coefficient $\rho
$ denotes the coupling constant between the polarization field $\hat X%
\left( x\right) $ and the electromagnetic field. It is worth noting that the
polarization field operator $\hat X\left( x,\omega \right) $ may be
expressed as a sum of the electric field operator $\hat E\left( x,\omega
\right) $ multiplied by $\left( \varepsilon \left( \omega \right) -1\right) $
and the bosonic operator $\hat f\left( x,\omega \right) $ multiplied by
the square root of the imaginary part of the complex permittivity, which
describes absorption in the dielectric. Explicit expression for the complex
permittivity $\varepsilon \left( \omega \right) $ can be found in \cite
{Huttner.Barnett.1992}, but these are not of relevance for our task, and any
phenomenologically introduced expression consistent with the Kramers-Kronig
relations can be used. Also the explicit expression of the reservoir field
operators $\hat Y\left( x,\omega \right) $ in terms of the bosonic
operators $\hat f\left( x,\omega \right) $ are not needed in the
following.

Using the commutation relations (\ref{com.rel.ff},\ref{com.rel.fff}) it can
be shown \cite{Huttner.Barnett.1992,Gruner.Welsch.1996} that the field
operators $\hat A\left( x\right) $ and $\hat E\left( x\right) $ and
the polarization field operators $\hat X\left( x\right) $ satisfy the
well known canonical commutation relations
\begin{equation}
\label{eq.comm.rel.A.E}\left[ \hat A\left( x\right) ,\hat E\left(
x^{\prime }\right) \right] =-\frac{i\hbar }{\epsilon _0{\cal A}}\delta
\left( x-x^{\prime }\right) ,
\end{equation}
\begin{equation}
\label{eq.comm.rel.X.E}\left[ \hat X\left( x\right) ,\hat E\left(
x^{\prime }\right) \right] =0,
\end{equation}
\begin{equation}
\label{eq.comm.rel.X.A}\left[ \hat X\left( x\right) ,\hat A\left(
x^{\prime }\right) \right] =0.
\end{equation}
Using the Green function (\ref{eq.GA}), Eq. (\ref{conn.Aw.int.dw.GA.f}) may
be written as
\begin{equation}
\label{eq.A.Apos.Aneg}\hat A\left( x,\omega \right) =\hat A%
_{(\rightarrow )}\left( x,\omega \right) +\hat A_{(\leftarrow )}\left(
x,\omega \right)
\end{equation}
where
\begin{equation}
\label{def.Apos}\hat A_{(\rightarrow )}\left( x,\omega \right)
=\int\limits_{-\infty }^xdx^{\prime }\ G_{{\rm A}}\left( x,x^{\prime
},\omega \right) \hat f\left( x^{\prime },\omega \right) =-i\alpha \sqrt{%
\frac{\varepsilon _{{\rm i}}}\varepsilon }\int\limits_{-\infty }^xdy\
e^{ik(x-y)}\hat f\left( y,\omega \right)
\end{equation}
and
\begin{equation}
\label{def.Aneg}\hat A_{(\leftarrow )}\left( x,\omega \right)
=\int\limits_x^\infty dx^{\prime }\ G_{{\rm A}}\left( x,x^{\prime },\omega
\right) \hat f\left( x^{\prime },\omega \right) =-i\alpha \sqrt{\frac{%
\varepsilon _{{\rm i}}}\varepsilon }\int\limits_x^\infty dy\ e^{-ik(x-y)}%
\hat f\left( y,\omega \right)
\end{equation}
denote vector potential operator components describing wave propagation in
the positive and negative x-direction respectively. From Eqs. (\ref{def.Apos}%
,\ref{def.Aneg}) we easily find
\begin{equation}
\label{eq.Langevin.pos}\partial _x\hat A_{(\rightarrow )}\left( x,\omega
\right) =ik\hat A_{(\rightarrow )}\left( x,\omega \right) -i\alpha \sqrt{%
\frac{\varepsilon _{{\rm i}}}\varepsilon }\hat f\left( x,\omega \right)
,
\end{equation}
\begin{equation}
\label{eq.Langevin.neg}\partial _x\hat A_{(\leftarrow )}\left( x,\omega
\right) =-ik\hat A_{(\leftarrow )}\left( x,\omega \right) +i\alpha \sqrt{%
\frac{\varepsilon _{{\rm i}}}\varepsilon }\hat f\left( x,\omega \right)
.
\end{equation}

These relations can be considered as quantum Langevin equations governing
the spatial evolution of the vector potential operator components $\hat A%
_{(\rightarrow )}\left( x,\omega \right) $ and $\hat A_{(\leftarrow
)}\left( x,\omega \right) $. Here the operators $\hat f\left( x,\omega
\right) $ play the role of the Langevin force operators. Note that these
noise terms are proportional to $\sqrt{\varepsilon _{{\rm i}}}$ which is
responsible for the absorption.

A similar decomposition as in Eq. (\ref{eq.A.Apos.Aneg}) can be performed
for the electric field operator components, where now the Green function $G_{%
{\rm E}}\left( x,x^{\prime },\omega \right) =i\omega G_{{\rm A}}\left(
x,x^{\prime },\omega \right) $ has to be used. Finally we consider the time
evolution of the operators in the Heisenberg picture. We immediately find
from Eqs.(\ref{def.HL.01},\ref{conn.HL.wff})
\begin{equation}
\label{eq.Heisenberg.f.lin}i\partial _t\hat f\left( x,\omega \right)
=\left[ \hat f\left( x,\omega \right) ,{\hat {{\cal H}}}_L\right]
=\omega \ \hat f\left( x,\omega \right)
\end{equation}
and similar equations for the other field operators, which are connected
with $\hat f\left( x,\omega \right) $. E.g. using Eq.(\ref{def.Apos}) we
have
\begin{equation}
\label{eq.Heisenberg.Apos.lin}i\partial _t\hat A_{(\rightarrow )}\left(
x,\omega \right) =\left[ \hat A_{(\rightarrow )}\left( x,\omega \right) ,%
{\hat {{\cal H}}}_L\right] =\omega \ \hat A_{(\rightarrow )}\left(
x,\omega \right) .
\end{equation}
It is seen from Eqs. (\ref{eq.Heisenberg.f.lin}, \ref{eq.Heisenberg.Apos.lin}%
) that the operators show a harmonic time evolution in the case of a linear
dielectric:
\begin{equation}
\label{eq.f.exp}\hat f\left( x,\omega ,t_0+t\right) =e^{-i\omega t}%
\hat f\left( x,\omega ,t_0\right) ,
\end{equation}
\begin{equation}
\label{eq.A.exp}\hat A_{(\rightarrow )}\left( x,\omega ,t_0+t\right)
=e^{-i\omega t}\hat A_{(\rightarrow )}\left( x,\omega ,t_0\right),
\end{equation}
and $\omega $ can be associated with the frequency of the harmonic time
evolution of all the operators with the argument $\omega $. Thus, in case of
a linear dielectric the decomposition in Eq.(\ref{conn.A.int.dw.A}) can be
considered as a Fourier decomposition of the operator of the vector
potential in the Heisenberg picture.

Concluding this section we state that the formulae given above enables one
to study quantum properties of light fields propagating in a linear
dielectric with absorption and dispersion \cite{Schmidt.1996} or passing
through a multi-layer structure \cite{Gruner.Welsch.io}.

\section{Nonlinear case}

\label{s3}

In this section we extend the linear model of a dielectric in order to
include non-linear optical effects. This offers the possibility for studying
the propagation of quantum light fields in nonlinear optical media including
dispersion and absorption. In the Hopfield model \cite{Hopf.1958} the
dielectric is described by a polarization field, the equations of motion for
this field being linear. In reality the material system is not composed of
harmonic oscillators and so this description is only an approximate one. In
order to introduce into the Hopfield-model the really existing
nonlinearities of the dielectric we add to the Lagrangian of the material
system an arbitrary functional $-\Phi [X\left( x\right) ]$ of the
polarization field $X\left( x\right) $. This term does not affect the
definition of the canonical momenta conjugated to the field variables.
Therefore the whole quantization scheme developed for the linear dielectric
can be used also in the case of the nonlinear dielectric. The Hamiltonian
describing the nonlinear dielectric interacting with the radiation field
then reads as
\begin{equation}
\label{def.HL.HN}{\bf \hat{H}}= \hbar \left( {\hat {{\cal H}}}_L+{%
\hat {{\cal H}}}_N\right) =\hbar \int\limits_{-\infty }^\infty dx\
\left( \hat{H}_L\left( x\right) +\hat{H}_N\left( x\right) \right) ,
\end{equation}
where $\hat{H}_L\left( x\right) $ is the linear part of the Hamiltonian
density which is defined in Eq.(\ref{conn.HL.wff}), and $\hat{H}_N\left(
x\right) $ describes the intrinsic nonlinearity of the dielectric,
\begin{equation}
\label{def.HDNL}\hat{H}_N\left( x\right) =\Phi [\hat{X}\left(
x\right) ].
\end{equation}
We emphasize that all the equations (\ref{conn.HL.wff}) -- (\ref
{eq.Langevin.neg}) between the electromagnetic field operators, the
polarization field, and the bosonic creation and destruction operators $%
\hat{f}^{\dagger }\left( x,\omega \right) $ and $\hat{f}\left(
x,\omega \right) $ are valid also in the case of the nonlinear dielectric
interacting with the electromagnetic field. These equations have been
derived in the case of the linear dielectric by the diagonalization of the
linear Hamiltonian \cite{Huttner.Barnett.1992,Gruner.Welsch.1996}, but
without using the Heisenberg equations of motion for the corresponding
operators. Therefore in the case of the nonlinear dielectric described by
the Hamiltonian (\ref{def.HL.HN},\ref{def.HDNL}), the same procedure can be
applied.
But with respect to the time development in the Heisenberg picture instead
of Eqs.(\ref{eq.Heisenberg.f.lin},\ref{eq.Heisenberg.Apos.lin}) now we find
\begin{equation}
\label{eq.Heisenberg.f.nl}i\partial _t\hat{f}\left( x,\omega \right)
=\left[ \hat{f}\left( x,\omega \right) ,{\hat {{\cal H}}}_L+{%
\hat {{\cal H}}}_N\right] =\omega \ \hat{f}\left( x,\omega \right)
+\left[ \hat{f}\left( x,\omega \right) ,{\hat {{\cal H}}}_N\right] ,
\end{equation}
\begin{equation}
\label{eq.Heisenberg.Apos.nl}i\partial _t\hat{A}_{(\rightarrow )}\left(
x,\omega \right) =\left[ \hat{A}_{(\rightarrow )}\left( x,\omega \right)
,{\hat {{\cal H}}}_L+{\hat {{\cal H}}}_N\right] =\omega \ \hat{A}%
_{(\rightarrow )}\left( x,\omega \right) +\left[ \hat{A}_{(\rightarrow
)}\left( x,\omega \right) ,{\hat {{\cal H}}}_N\right] .
\end{equation}
These equations of motion, together with the quantum Langevin equations (\ref
{eq.Langevin.pos},\ref{eq.Langevin.neg}) governing the spatial evolution of
the vector potential operator components describe the space time behavior of
the radiation field in a non-linear, dispersive and absorbing dielectric.
These equations have to be supplemented by the relation (\ref{conn.Xw.Ew.f})
between the polarization field operator $\hat{X}\left( x,\omega \right) $%
, the operator of the electric field $\hat{E}\left( x,\omega \right) $
and the bosonic operator $\hat{f}\left( x,\omega \right) $. Using this
equation we can express the non-linear part of the Hamiltonian completely in
terms of $\hat{E}\left( x,\omega \right) $ and $\hat{f}\left(
x,\omega \right) $ and their Hermitian conjugate operators. As a consequence
Eqs.(\ref{eq.Heisenberg.f.nl},\ref{eq.Heisenberg.Apos.nl}) include nonlinear
optical effects in the time development of the radiation field.
As can be seen from the Heisenberg equations of motion (\ref
{eq.Heisenberg.f.nl},\ref{eq.Heisenberg.Apos.nl}) due to the nonlinear part
of the Hamiltonian the time development of the corresponding operators is
much more complicated as in the linear case. Therefore the decomposition Eq.(%
\ref{conn.A.int.dw.A}) now cannot be considered as a Fourier decomposition
of the operator of the vector potential in the Heisenberg picture, because
these components show a very complicated time behavior, and the parameter $%
\omega $ cannot be associated with the frequency in the Fourier space.

Finally we would like to note that the diagonalization procedure remains the
same also in case of a more general nonlinear part $\hat{H}_N\left(
x\right) $ of the Hamiltonian, which may be a functional of the operators $%
\hat{A}\left( x\right) $, $\hat{X}\left( x\right) $, $\hat{Y}%
\left( x,\omega \right) $ and its spacial derivatives
(see e.g. \cite{Hillery.1996}), and the Eqs. (\ref
{conn.HL.wff}) -- (\ref{eq.Langevin.neg}) as well as the Heisenberg
equations of motion (\ref{eq.Heisenberg.f.nl},\ref{eq.Heisenberg.Apos.nl})
may be used in order to investigate the radiation field inside the nonlinear
dielectric. This offers the possibility to
generalize the Hopfield model if necessary.

\section{Pulse propagation equation.}

\label{s4}

In order to study the propagation of quantum pulses in nonlinear dispersive
and absorptive dielectrics we have to solve the nonlinear propagation
equations in time (\ref{eq.Heisenberg.f.nl},\ref{eq.Heisenberg.Apos.nl})
together with the quantum Langevin equations (\ref{eq.Langevin.pos},\ref
{eq.Langevin.neg}) governing the spatial evolution. In this section we shall
consider these equations in more detail in the case of narrow-frequency band
quantum pulses, and eventually for a Kerr medium we derive the quantum
version of the soliton equation including also the effect of absorption.

To begin with we divide the $\omega $ -interval $\left[ 0..\infty \right] $
into subintervals and decompose the field operators as follows:
\begin{equation}
\label{def.A.sum.Ai}\hat{A}^{(+)}\left( x\right) =\sum_{i=0}^\infty
\hat{A}_i^{(+)}\left( x\right) ,
\end{equation}
\begin{equation}
\label{def.Ai}\hat{A}_i^{(+)}\left( x\right) =\int\limits_{-\Delta
\omega /2}^{\Delta \omega /2}d\Omega \ \hat{A}\left( x,\omega ^i+\Omega
\right) ,
\end{equation}
where
\begin{equation}
\label{def.wi}\omega ^i=\left( i+1/2\right) \Delta \omega .
\end{equation}
The analogous relations can be written for all the other operators which can
be expressed by an integral over $\omega $. Now we assume that the light
pulse under consideration has a small spectral bandwidth, so that only
components of the field operator in one of the intervals, characterized by
an index $i_0$, must be considered. It means that the bandwidth should not
exceed the length of the corresponding subinterval $\Delta \omega $. This
also gives a criterium for choosing the value of $\Delta \omega $. In the
following we restrict the parameter $\omega $ to this interval, and we write
\begin{equation}
\label{conn.w.w0.W}\omega =\omega _0+\Omega
\end{equation}
where $\omega _0=\omega ^{i_0}$. Next we introduce slowly varying operators $%
\hat{\underline{A}}_{(\rightarrow )}\left( x,\Omega \right) $ (operators
corresponding to the other direction can be introduced in a similar way) and
$\hat{\underline{f}}_{\left( \rightarrow \right) }\left( x,\Omega
\right) $ according to
\begin{eqnarray}
&  & \hat A_{(\rightarrow )}\left( x,\omega \right) =e^{i\varphi
_{(\rightarrow )}}\hat{\underline{A}}_{(\rightarrow )}\left( x,\Omega
\right), \nonumber \\  &  & \hat{f}\left( x,\omega \right) =e^{i\varphi
_{(\rightarrow )}}\hat{\underline{f}}_{\left( \rightarrow \right)
}\left( x,\Omega \right)
\label{def.SVA}
\end{eqnarray}
(the index $i_0$ will be suppressed here and in the following), where
\begin{equation}
\label{def varphi}\varphi _{(\rightarrow )}=k_\varphi x-\omega _0t,
\end{equation}
and $k_\varphi $ is a real number
to be determined later.

Using the Heisenberg equations of motion (\ref{eq.Heisenberg.f.nl}, \ref
{eq.Heisenberg.Apos.nl}) we find
\begin{equation}
\label{eq.WA}\Omega \ \hat {\underline{A}}_{(\rightarrow )}\left(
x,\Omega \right) =\left( i\partial _t+{\hat {{\cal H}}}_N^{\times
}\right) \hat {\underline{A}}_{(\rightarrow )}\left( x,\Omega \right) ,
\end{equation}
\begin{equation}
\label{eq.Wf}\Omega \ \hat {\underline{f}}_{\left( \rightarrow \right)
}\left( x,\Omega \right) =\left( i\partial _t+{\hat {{\cal H}}}%
_N^{\times }\right) \hat {\underline{f}}_{\left( \rightarrow \right)
}\left( x,\Omega \right) ,
\end{equation}
where the notation
\begin{equation}
\label{def.H.x}{\hat {{\cal H}}}_N^{\times }\ \hat O\equiv \left[ {%
\hat {{\cal H}}}_N,\hat O\right]
\end{equation}
has been introduced. We rewrite also the quantum Langevin equation (\ref
{eq.Langevin.pos}) in terms of the new slowly varying operators (Eqs. (\ref
{def.SVA})):
\begin{equation}
\label{eq.Langevin.pos.SVA}\partial _x\hat {\underline{A}}_{(\rightarrow
)}\left( x,\Omega \right) =i\left( k-k_\varphi \right) \hat {\underline{A%
}}_{(\rightarrow )}\left( x,\Omega \right) -i\alpha \sqrt{\frac{\varepsilon
_{{\rm i}}}\varepsilon }\ \hat {\underline{f}}_{\left( \rightarrow
\right) }\left( x,\Omega \right) .
\end{equation}
To be able to treat these equations (\ref{eq.WA},\ref{eq.Wf},\ref
{eq.Langevin.pos.SVA}) in more detail we expand the wave number $k$ as a
function of $\omega $ near $\omega _0$ into a Taylor series
\begin{equation}
\label{eq.k.serie}k=\sum\limits_{m=0}^\infty k_m\frac{\Omega ^m}{m!},
\end{equation}
and in a similar way we have
\begin{equation}
\label{eq.eps.serie}\sqrt{\frac{\varepsilon _{{\rm i}}}\varepsilon }%
=\sum\limits_{m=0}^\infty p_m\frac{\Omega ^m}{m!},
\end{equation}
where the coefficients $k_m\equiv k_{mr}+ik_{mi}$ and $p_m\equiv
p_{mr}+ip_{mi}$ are in general complex. Substituting these expansions into
Eq.(\ref{eq.Langevin.pos.SVA}) and employing Eqs.(\ref{eq.WA},\ref{eq.Wf}),
we arrive at
\begin{eqnarray} \label{eq.Langevin.pos.HN}
&  & \partial _x
\hat {\underline{A}}_{(\rightarrow )}\left( x,\Omega \right) =i\left(
\sum\limits_{m=0}^\infty \frac{k_m}{m!}\left( i\partial _t+{\hat {{\cal H%
}}}_N^{\times }\right) ^m-k_\varphi \right) \hat{\underline{A}}%
_{(\rightarrow )}\left( x,\Omega \right) \nonumber \\  &  & \qquad -i\alpha
\sum\limits_{m=0}^\infty \frac{p_m}{m!}\left( i\partial _t+{\hat {{\cal H%
}}}_N^{\times }\right) ^m\ \hat{\underline{f}}_{\left( \rightarrow
\right) }\left( x,\Omega \right) .\
\end{eqnarray}
Integration over $\Omega $ gives
\begin{equation}
\label{eq.prop.01}\partial _x\hat {\underline{A}}_{(\rightarrow
)}^{(+)}\left( x\right) =i\left( \sum\limits_{m=0}^\infty \frac{k_m}{m!}%
\left( i\partial _t+{\hat {{\cal H}}}_N^{\times }\right) ^m-k_\varphi
\right) \hat {\underline{A}}_{(\rightarrow )}^{(+)}\left( x\right) +%
\mbox{R.S.},
\end{equation}
where $\hat {\underline{A}}_{(\rightarrow )}^{(+)}\left( x\right) $ is
given by the relations
\begin{eqnarray} \label{eq.prop.02}
&  & \hat A^{(+)}\left( x\right) =\hat A_{(\rightarrow )}^{(+)}\left(
x\right) +\hat{A}_{(\leftarrow )}^{(+)}\left( x\right) ,\nonumber \\  &
& \hat{A}_{(\rightarrow )}^{(+)}\left( x\right) =e^{i\varphi
_{(\rightarrow )}}\hat{\underline{A}}_{(\rightarrow )}^{(+)}\left(
x\right) ,
\end{eqnarray}
and
\begin{eqnarray} \label{eq.prop.021}
&  & \mbox{R.S.}=-i\alpha \sum\limits_{m=0}^\infty \frac{p_m}{m!}\left(
i\partial _t+ {\hat {{\cal H}}}_N^{\times }\right)
^m\int\limits_{-\Delta \omega /2}^{\Delta \omega /2}d\Omega \ \hat{%
\underline{f}}_{\left( \rightarrow \right) }\left( x,\Omega \right) .
\end{eqnarray}
Note that the various operators $\hat A$ and $\hat{f}$ should be
labeled by the index $i_0$, describing the parameter range to which these
operators belong (see e.g. Eq.(\ref{def.A.sum.Ai})). The equations (\ref
{eq.prop.01} -- \ref{eq.prop.021}) describe the propagation of a quantum
pulse with a narrow bandwidth spectrum in the dielectric with any order of
dispersion, absorption and nonlinearity.
In general we need similar equations for the fields running in the other
direction because they are included in ${\hat {{\cal H}}}_N$. It should
be mentioned that the restriction to a pulse with narrow spectral bandwidth
is not necessary, but in the general case we must consider for each $\omega $
-interval such an equation as Eq.(\ref{eq.prop.01}).

Let us consider a special case with small dispersion, where the absorption
in the frequency interval around $\omega _{i_{0}}$ can be neglected ($%
\varepsilon _{i}\left( \omega \right) =$Im$(\varepsilon \left( \omega
\right) )=0$, $k_{mi}=0$ , $k_{1r}=1/v_{gr}$, where $v_{gr}$ is the group
velocity, and $k_{mr}=0$ for $m\ge 2$). Choosing $k_{\varphi }=k_{0r}$ we
arrive at the simplest case of a propagation equation for a quantum light
pulse in a nonlinear dielectric,
\begin{equation}
\label{eq.prop.03}\left[ \partial _{x}+\frac{1}{v_{gr}}\partial _{t}\right]
\hat{\underline{A}}_{(\rightarrow )}^{(+)}\left( x\right) =\frac{i}{c}%
\left[ {\hat{{\cal H}}}_{N},\hat{\underline{A}}_{(\rightarrow
)}^{(+)}\left( x\right) \right] .
\end{equation}
Note that the assumption of zero absorption resulted in the vanishing of the
term proportional to the operator $\hat{\underline{f}}_{\left(
\rightarrow \right) }\left( x,\Omega \right) $, which may be considered as a
Langevin fluctuation operator. Such operators have to be included in the
Heisenberg equation of motion in order to describe correctly also losses.

\subsection{Second order dispersion approximation.}

\label{s4.1}

Now we consider the propagation of a quantum pulse with narrow spectral
bandwidth including to some extend dispersion and absorption. We neglect all
terms proportional to $\Omega ^{m}$ with $m>2$. That means we take into
account the dispersion only up to the second order. Moreover we neglect all
terms in which the nonlinear Hamiltonian ${\hat{{\cal H}}}_{N}$ and the
Langevin fluctuation operators $\hat{\underline{f}}_{\left( \rightarrow
\right) }\left( x,\Omega \right) $ are coupled with higher order dispersion
and absorption effects (this means terms proportional to $k_{m}{\hat{%
{\cal H}}}_{N}^{\times }$, $m\ge 2$ and $p_{m}\hat{\underline{f}}%
_{\left( \rightarrow \right) }\left( x,\Omega \right) $, $m\ge 1$). Choosing
$k_{\varphi }=k_{0r}$, from Eq.(\ref{eq.prop.01}) we arrive at the following
propagation equation for a quantum light pulse
\begin{eqnarray}
&  & \left( i\partial _{x}+ik_{0
{\rm i}}+ik_{1}\partial _{t}-1/2k_{2}\partial _{tt}+k_{1}{\hat{{\cal H}}}%
_{N}^{\times }\right) \hat{\underline{A}}_{(\rightarrow )}^{(+)}\left(
x\right) \nonumber \\  &  & \qquad =\alpha p_{0}\int\limits_{-\Delta \omega
/2}^{\Delta \omega /2}d\Omega \ \hat{\underline{f}}_{(\rightarrow
)}\left( x,\Omega \right) , \label{eq.SOD.02}
\end{eqnarray}
which will be discussed in connection with a Kerr medium in the next
subsection.

\subsection{Approximation for the propagation in a Kerr medium.}

\label{s4.2}

In this section as an application of Eq.(\ref{eq.SOD.02}) we discuss the
pulse propagation in a Kerr medium. At first we specify the nonlinear part
of the Hamiltonian ${\hat{{\cal H}}}_N$. The most simple extension of
the linear dielectric to include nonlinearities is in case of a nonlinear
dielectric with inversion symmetry the Ansatz
\begin{equation}
\label{def.HNPHI.kerr}\Phi [\hat{X}\left( x\right) ]=\frac \lambda {4!}%
\hat{X}^4\left( x\right) ,
\end{equation}
from which we find (see Eq.(\ref{def.HDNL}))
\begin{equation}
\label{def.HN.kerr}{\hat{{\cal H}}}_N=\frac \lambda {4!}%
\int\limits_{-\infty }^\infty dx^{\prime }\hat{X}^4\left( x^{\prime
}\right) ,
\end{equation}
where $\lambda $ is a constant characterizing the nonlinearity in a
homogeneous dielectric. In considering the propagation of a quantum pulse
with narrow spectral bandwidth into the positive x-direction (see Eq. (\ref
{eq.SOD.02})) we may rewrite ${\hat{{\cal H}}}_N$ approximately as
\begin{equation}
\label{def.HN.kerr.01}{\hat{{\cal H}}}_N=\frac \lambda 4%
\int\limits_{-\infty }^\infty dx^{\prime }\ \hat{\underline{X}}%
_{(\rightarrow )}^{(-)}\left( x^{\prime }\right) \ \hat{\underline{X}}%
_{(\rightarrow )}^{(-)}\left( x^{\prime }\right) \ \hat{\underline{X}}%
_{(\rightarrow )}^{(+)}\left( x^{\prime }\right) \ \hat{\underline{X}}%
_{(\rightarrow )}^{(+)}\left( x^{\prime }\right) ,
\end{equation}
where (remember Eq.(\ref{conn.w.w0.W}))
\begin{eqnarray}
&  & \hat {\underline{X}}_{(\rightarrow )}^{(+)}\left( x^{\prime }\right)
=\int\limits_{-\Delta \omega /2}^{\Delta \omega /2}d\Omega \left[ i\omega
\left\{ \varepsilon \left( \omega \right) -1\right\} \hat{\underline{A}}%
_{(\rightarrow )}\left( x,\Omega \right) \right. \nonumber \\  &  & \qquad
\left. -2i\alpha c\sqrt{\varepsilon _{{\rm i}}\left( \omega \right) }%
\hat{\underline{f}}_{(\rightarrow )}\left( x,\Omega \right) \right] ,%
\label{def.HN.kerr.011}
\end{eqnarray}
and the operators $\hat{\underline{A}}_{(\rightarrow )}\left( x,\Omega
\right) $ and $\hat{\underline{f}}_{(\rightarrow )}\left( x,\Omega
\right) $ are defined in Eq.(\ref{def.SVA}). Here we have also introduced a
normal ordering of the operators and quadratic terms in $\hat{X}$ have
been neglected. Straightforward calculations give for the commutator term $%
\left[ {\hat{{\cal H}}}_N,\hat{\underline{A}}_{(\rightarrow
)}^{(+)}\left( x\right) \right] $ of Eq. (\ref{eq.SOD.02})
\begin{equation}
\label{eq.HNA.07}\left[ {\hat{{\cal H}}}_N,\hat{\underline{A}}%
_{(\rightarrow )}^{(+)}\left( x\right) \right] =\frac \lambda 2%
\int\limits_{-\infty }^\infty dx^{\prime }\ G_H\left( x,x^{\prime }\right)
\hat{\underline{X}}_{(\rightarrow )}^{(-)}\left( x^{\prime }\right) \
\hat{\underline{X}}_{(\rightarrow )}^{(+)}\left( x^{\prime }\right) \
\hat{\underline{X}}_{(\rightarrow )}^{(+)}\left( x^{\prime }\right) ,
\end{equation}
where
\begin{eqnarray}
&  & G_H\left( x,x^{\prime }\right) =\alpha ^2
\frac{\epsilon _0}\rho \int\limits_{-\Delta \Omega /2}^{\Delta \omega
/2}d\Omega \left( i\omega \left( \varepsilon ^{*}-1\right) \frac{\varepsilon
_{{\rm i}}}{2k_{{\rm i}}\left| \varepsilon \right| }-2c\frac{\varepsilon _{%
{\rm i}}}{\sqrt{\varepsilon }}U\left( \Delta x\right) \right) \nonumber \\
&  & \qquad \qquad \times \exp \left[ -k_{{\rm i}}\left| \Delta x\right|
+i\left( k_{{\rm r}}-k_{0{\rm r}}\right) \Delta x\right] ,\label{eq.HNA.08}
\end{eqnarray}
$U\left( \Delta x\right) $ denotes the unity step function, and
$\Delta x=x-x^{\prime }$.
We note that the relation (\ref{eq.HNA.07}) in general describes a nonlocal,
nonlinear response of the matter, and Eq. (\ref{eq.SOD.02}) is an
integro-differential equation with respect to the space co-ordinate $x$.

Now we consider the limiting case of weak absorption in the interval around $%
\omega _0$. Employing the approximation
\begin{equation}
\label{epsi.epsr}\varepsilon _{{\rm i}}\ll \varepsilon _{{\rm r}},
\end{equation}
the term $\frac{\varepsilon _{{\rm i}}}{2k_{{\rm i}}\left| \varepsilon
\right| }$ may be approximated by
\begin{equation}
\label{rat.01}\frac{\varepsilon _{{\rm i}}}{2k_{{\rm i}}\left| \varepsilon
\right| }\sim \frac{\varepsilon _{{\rm i}}}{2\frac \omega c\sqrt{\varepsilon
_{{\rm r}}}\,\mbox{{\rm Im}}\sqrt{1+i\frac{\varepsilon _{{\rm i}}}{%
\varepsilon _{{\rm r}}}}\left| \varepsilon \right| }\sim \frac{\varepsilon _{%
{\rm i}}}{2\frac \omega c\sqrt{\varepsilon _{{\rm r}}}\,\mbox{{\rm Im}}%
\!\left( 1+i\frac{\varepsilon _{{\rm i}}}{2\varepsilon _{{\rm r}}}\right)
\varepsilon _{{\rm r}}}\sim \frac 1{k_{{\rm r}}},
\end{equation}
and the term containing the step function can be neglected, because it is
proportional to $\varepsilon _{{\rm i}}$.
As a further approximation we expand the integrand of the integral kernel in
powers of $\Omega $ and retain only the lowest order terms arriving at
\begin{eqnarray}
&  & G_H\left( x,x^{\prime }\right) =\alpha ^2
\frac{\epsilon _0}\rho i\omega _0\left( \varepsilon ^{*}-1\right) \frac 1{k_{%
{\rm r}}}\int\limits_{-\Delta \omega /2}^{\Delta \omega /2}d\Omega \exp
\left[ i\Omega \ \Delta x\ k_{1{\rm r}}\right] ,\nonumber \\  &  & \qquad
=\alpha ^2\frac{\epsilon _0}\rho i\omega _0\left( \varepsilon ^{*}-1\right)
\frac 2{k_{{\rm r}}\ \Delta x\ k_{1{\rm r}}}\sin \left[ \Delta \omega \ k_{1%
{\rm r}}\ \Delta x/2\right] .\label{def.GH}
\end{eqnarray}
Assuming that the operators $\hat{\underline{X}}_{(\rightarrow )}^{(+)}$
in Eq.(\ref{eq.HNA.07}) are slowly varying on a scale defined by the
characteristic length $\left( \Delta \omega \ k_{1{\rm r}}\right) ^{-1}$ of $%
G_H\left( x,x^{\prime }\right) $, we can replace the function $G_H\left(
x,x^{\prime }\right) $ by a delta-function,
and the commutator term in Eq.(\ref{eq.HNA.07}) reads as
\begin{equation}
\label{eq.k1.HNA}\left[ \hat{{\cal H}}_N,\hat{\underline{A}}%
_{(\rightarrow )}^{(+)}\left( x\right) \right] =\frac{\pi \alpha ^2\lambda
\epsilon _0}{k_{{\rm r}}k_{1{\rm r}}\rho }i\omega _0\left( \varepsilon
^{*}-1\right) \ \hat{\underline{X}}_{(\rightarrow )}^{(-)}\left(
x\right) \ \hat{\underline{X}}_{(\rightarrow )}^{(+)}\left( x\right) \
\hat{\underline{X}}_{(\rightarrow )}^{(+)}\left( x\right) .
\end{equation}
Higher order approximations
in Eq. (\ref{eq.HNA.08}) give rise to nonlocal corrections in Eq. (\ref
{eq.HNA.07}), which will be neglected here.
In the same way as we have arrived at the
Eq.(\ref{eq.SOD.02}), we neglect in Eq.(\ref{eq.k1.HNA}) the Langevin
fluctuation operators $\hat{\underline{f}}_{\left( \rightarrow \right)
}\left( x,\Omega \right) $, which result from the decomposition of $\hat{%
\underline{X}}_{(\rightarrow )}^{(+)}\left( x\right) $ (see Eq.(\ref
{def.HN.kerr.011})), and we substitute $k_{1{\rm r}}$ for $k_1$, thus
finding eventually
\begin{equation}
\label{def.HN.kerr.012}k_{1{\rm r}}\left[ \hat{{\cal H}}_N,\hat{%
\underline{A}}_{(\rightarrow )}^{(+)}\left( x\right) \right] =\chi \
\hat{\underline{A}}_{(\rightarrow )}^{(-)}\left( x\right) \ \hat{%
\underline{A}}_{(\rightarrow )}^{(+)}\left( x\right) \ \hat{\underline{A}%
}_{(\rightarrow )}^{(+)}\left( x\right) ,
\end{equation}
where
\begin{equation}
\label{def.chi.eff}\chi =\frac{\pi \alpha ^2}{k_{{\rm r}}}\lambda \left(
\frac{\epsilon _0}\rho \omega _0\left| \varepsilon -1\right| \right) ^4.
\end{equation}
Substituting this in Eq. (\ref{eq.SOD.02})
we obtain finally the following equation for the propagation of a
narrow-frequency band quantum pulse in a Kerr medium with dispersion and
weak absorption
\begin{eqnarray}
&  & \left( i\partial _x+ik_{0
{\rm i}}+ik_1\partial _t-1/2k_2\partial _{tt}+\chi \ \hat{\underline{A}}%
_{(\rightarrow )}^{(-)}\left( x\right) \hat{\underline{A}}_{(\rightarrow
)}^{(+)}\left( x\right) \right) \hat{\underline{A}}_{(\rightarrow
)}^{(+)}\left( x\right) \nonumber \\  &  & \qquad =\alpha \sqrt{\frac{%
\varepsilon _{{\rm i}}}\varepsilon }\int\limits_{-\Delta \omega /2}^{\Delta
\omega /2}d\Omega \ \hat{\underline{f}}_{(\rightarrow )}\left( x,\Omega
\right) .\label{eq.schrod}
\end{eqnarray}
The absorption properties of the dielectric are described by the terms
proportional to $k_{m{\rm i}}$ and the Langevin fluctuation operator on the
right hand side of this equation. The term proportional to $\chi $ describes
the Kerr nonlinearity, and the dispersion effects are contained in the term
proportional to $k_{2{\rm r}}$. Of course the solution of this nonlinear
operator equation containing also a Langevin fluctuation operator will be
difficult, and numerical methods must be employed.

It is clearly seen that in the case of pure real coefficients $k_1$ and $k_2$
in Eq.(\ref{eq.schrod}) and vanishing $\varepsilon _{{\rm i}}$ we obtain the
operator version of the classical (see e.g. \cite
{Hasegawa.1989,Akhmanov.1992}) Schr\"odinger equation similar to the one
used in \cite{Lai.Haus.1989a,Lai.Haus.1989b}. We note that such an
approximation is in general not consistent with the Kramers-Kronig relations
because dispersion is always accompanied by absorption.
Including the absorption brings an additional fluctuation term into the Eq.(%
\ref{eq.schrod}) which is important, in order that the canonical field
commutation relations Eq.(\ref{eq.comm.rel.A.E}) are preserved.

\section{Summary}

\label{s5}

On the basis of the quantization of the radiation field in a linear
dielectric with dispersion and absorption as it has been described by
Huttner and Barnett \cite{Huttner.Barnett.1992} and Gruner and Welsch \cite
{Gruner.Welsch.1996}, we have developed a quantization scheme which includes
besides linear dispersion and absorption also nonlinear optical effects.
This has been achieved by introducing into the matter part of the
microscopic Huttner-Barnett model a nonlinear interaction term depending
only on the polarization field operator which describes the dielectric. This
guaranties, that the operator relations at equal times of the linear theory
are also valid in the extended theory. Especially the equal time commutation
relations of the free field case are preserved, and the space propagation
equations for field components are the same as in the case of a linear
dielectric. We have applied the theory to a narrow-frequency band quantum
pulse propagating in a dielectric with a Kerr-like nonlinearity. Using a
second order dispersion approximation and assuming weak absorption in the
frequency band under consideration we have derived in the Heisenberg picture
a nonlinear operator equation which governs the space-time development of
the quantum pulse. It contains the Kerr nonlinearity, second order
dispersion and an absorption term together with a corresponding fluctuation
operator. The resulting equation can be considered as a generalization of a
nonlinear Schr\"odinger equation describing quantum solitons.

\section{Acknowledgments}

We are grateful to T. Gruner for valuable discussions. This work was
supported by the Deutsche Forschungsgemeinschaft (74021\thinspace
40\thinspace 223).

%\end{thebibliography}


\begin{references}
\bibitem{Drummond.Nature.1993}  P.D. Drummond, R.M. Shelby, S.R. Friberg,
and Y. Yamamoto, Nature {\bf 365}, 307 (1993),

\bibitem{Hasegawa.1989}  A. Hasegawa, {\em Optical Solitons in Fibers}
(Springer-Verlag Berlin Heidelberg 1989).

\bibitem{Akhmanov.1992}  S.A. Akhmanov, V.A. Vysloukh, and A.S. Chirkin,
{\em Optics of Femtosecond Laser Pulses} (American Institute of Physics, New
York, 1992).

\bibitem{Hopf.1958}  J. J. Hopfield, Phys. Rev. {\bf 112}, 1555 (1958).

\bibitem{Pantell.1969}  E. H. Pantell and H. E. Puthoff, {\em Fundamentals
of Quantum Electronics} (Wiley, New York, 1969).

\bibitem{Marcuse.1980}  D. Marcuse, {\em Principles of Quantum Electronics}
(Academic, New York, 1980), chaps. 2 and 3.

\bibitem{Yariv.1989}  A. Yariv, {\em Quantum Electronics} 3rd ed. (Wiley,
New York, 1989).

\bibitem{Abram.1987}  J. Abram, Phys. Rev. A {\bf 35}, 4661 (1987).

\bibitem{Kn.Vo.We.1987}  L. Kn\"oll, W. Vogel, and D.-G. Welsch, Phys. Rev.
A {\bf 36}, 3803 (1987); W. Vogel and D.-G. Welsch, {\em Lectures on Quantum
Optics} (Akademie Verlag, Berlin/VCH Publishers, New York, 1994).

\bibitem{Glauber.Lewenstein.1991}  R. J. Glauber and M. Lewenstein, Phys.
Rev. A {\bf 43}, 467 (1991).

\bibitem{Hillery.1984}  D. M. Hillery and L. D. Mlodinov, Phys. Rev. A {\bf %
30}, 1860 (1984).

\bibitem{Drummond.1987}  P. D. Drummond and S. J. Carter, J. Opt. Soc. Am. B
{\bf 4}, 1565 (1987).

\bibitem{Lai.Haus.1989a}  Y. Lai and H. A. Haus, Phys. Rev. A {\bf 40}, 844
(1989).

\bibitem{Lai.Haus.1989b}  Y. Lai and H. A. Haus, Phys. Rev. A {\bf 40}, 854
(1989).

\bibitem{Drummond.1990}  P. D. Drummond, Phys. Rev. A {\bf 42}, 6845 (1990).

\bibitem{Huttner.Barnett.1992}  B. Huttner and S.M. Barnett, Europhys. Lett.
{\bf 18}, 487 (1992); Phys. Rev. A {\bf 46} 4306 (1992).

\bibitem{Ho.Kumar.1993}  S. T. Ho and P. Kumar, J. Opt. Soc. Am. B {\bf 10},
1620 (1993).

\bibitem{Gruner.Welsch.1995}  T. Gruner and D.-G. Welsch,
% Corr Funct of...
Phys. Rev. A {\bf 51}, 3246 (1995).

\bibitem{Matloob.1995}  R. Matloob, R. Loudon, S. M. Barnett, and J.
Jeffers, Phys. Rev. A {\bf 52}, 4823 (1995).

\bibitem{Gruner.Welsch.1996}  T. Gruner and D.-G. Welsch,
% Preprint FSUJ TPI QO-01/95 (March, 1995);
% Green Function Appr ...
Phys. Rev. A {\bf 53}, 1818 (1996).

\bibitem{Hillery.1996}  D. M. Hillery, Acta. Phys. Slov. {\bf 46}, 259
(1996).

\bibitem{Carter.1995}  S. J. Carter, Phys. Rev. A {\bf 51}, 3274 (1995).

\bibitem{Gruner.Welsch.io}  T. Gruner and D.-G. Welsch,
% Input Output Relat....
Preprint FSUJ TPI QO-06/95 (October, 1995); to be published in Phys. Rev. A.

\bibitem{Schmidt.1996}  E. Schmidt, L. Kn\"oll and D.-G. Welsch,
%Preprint FSUJ TPI QO-10/95 (December, 1995);
Phys. Rev. A, in press.
\end{references}
\end{document}